\documentclass[aps,preprint,a4paper,amsmath,showpacs,floatfix]{revtex4}

\usepackage{graphicx}
\usepackage{epsfig}
\usepackage{amssymb}
\usepackage{times}

\begin{document}

\title{Phase behavior of parallel hard cylinders}

\author{Jos\'e A. Capit\'an, Yuri Mart\'{\i}nez-Rat\'on, and Jos\'e A. Cuesta}

\affiliation{Grupo Interdisciplinar de Sistemas Complejos (GISC),
Departamento de Matem\'aticas, Escuela Polit\'ecnica Superior,
Universidad Carlos III de Madrid,
Avenida de la Universidad 30, E-28911 Legan\'es, Madrid, Spain}

\date{\today}

\begin{abstract}
We test the performance of a recently proposed
fundamental measure density functional of aligned hard cylinders by
calculating the phase diagram of a monodisperse fluid of these particles.
We consider all possible liquid crystalline symmetries, namely nematic,
smectic and columnar, as well as the crystalline phase. For this purpose
we introduce a Gaussian parameterization of the density profile and use
it to minimize numerically the functional. We also determine, from the
analytic expression for the structure factor of the uniform fluid, the
bifurcation points from the nematic to the smectic and columnar phases.
The equation of state, as obtained from functional minimization, is
compared to the available Monte Carlo simulation. The agreement is
is very good, nearly perfect in the description of the inhomogeneous
phases. The columnar phase is found to be metastable with respect to
the smectic or crystal phases, its free energy though being very close
to that of the stable phases. This result justifies the observation of
a window of stability of the columnar phase in some simulations, which
disappears as the size of the system increases. The only important 
deviation between theory and simulations shows up in the location of
the nematic-smectic transition. This is the common drawback of any
fundamental measure functional of describing the uniform phase just 
with the accuracy of scaled particle theory.
\end{abstract}

\pacs{64.70.Md, 61.20.Gy, 05.20.Jj}

\maketitle

\section{Introduction}

Monte Carlo simulations  conducted on systems of hard anisotropic 
particles (spherocylinders being the most paradigmatic shape) 
showed that the purely entropic nature of hard core interactions is enough 
to explain the stability of different liquid-crystalline phases and 
phase transitions between them \cite{frenkel:1987a,frenkel:1987b,frenkel:1988}.
These phases, in decreasing order of symmetry, are known as isotropic (I),
nematic (N), smectic-A (Sm), columnar (C) and crystal (K) ---the isotropic
and the crystal not being liquid crystalline phases properly speaking---,
and some of their physical and chemical  properties have been described in detail
in Refs.~\cite{degennes:1994,chandrasekhar:1992}. Later, Monte Carlo simulations 
were also employed to calculate the full phase diagram of fluids of
freely-rotating hard spherocylinders \cite{bolhuis:1997}
and hard-cut spheres \cite{veerman:1992}, including non-uniform 
phases as the periodic one-dimensional (Sm), two-dimensional (C) and
three-dimensional (K) phases. 

Several density functional theories have been devised to determine
the phase behavior of the hard sphere (HS) fluid. These theories can 
be grouped in two different sets. The first one, the weighted-density functionals,
are constructed from the knowledge of the thermodynamical and structural
properties of the uniform fluid \cite{tarazona:1984,tarazona:1985,curtin:1985},
while the second one, the fundamental measure functionals (FMF), initially
introduced by Rosenfeld \cite{rosenfeld:1989,rosenfeld:1990}
and later improved for an adequate description of the HS freezing
\cite{rosenfeld:1996,rosenfeld:1997,tarazona:1997}, are built on
the geometry of the particles alone. 

The extensions of these theories to hard anisotropic particles have not
been as successful as they have been for HS. There are two reasons to explain this 
difficulty: The first one is related to the, as of today, still poor knowledge of
the structural properties of fluids composed by anisotropic particles, and the
second one is the inherent complexity in dealing with orientational degrees of
freedom within density functional theory. This notwithstanding, some
weighted-density functionals have been developed for the fluid of hard
spherocylinders \cite{poniewierski:1988,somoza:1989a} to study
both the I-Sm and the N-Sm phase transitions as a function of the particle aspect
ratio. These functionals were constructed as modifications of a reference HS
weighted-density functionals, and their predictions,
tested against Monte Carlo simulations, are reasonably good. They do not allow though
to properly account for the C and K phases.

FMF are more appropriate to treat these phases as, by construction, they are
more suitable to describe highly confined particles, such as they are in a
solid. Unfortunately the fundamental measure formalism has little flexibility
to apply it to arbitrary geometries. FMF have been obtained for parallelepipeds with
restricted orientations of their principal axes \cite{cuesta:1996,cuesta:1997a,cuesta:1997b},
and very recently for cylinders also with a parallel alignment constraint
\cite{martinez-raton:2008}. For freely rotating anisotropic particles FMF have
been obtained for needles, infinitely thin plates, and their mixtures
\cite{schmidt:2001a,brader:2002,esztermann:2004,esztermann:2006}, but this time
the price to pay is to eliminate at least one of the characteristic lengths of
the particles. Besides, the numerical minimization of these functionals to obtain
the equilibrium density profiles of non-uniform phases seems to be a very demanding
task.    
 
In this article we aim at testing the recently proposed FMF for parallel hard cylinders
\cite{martinez-raton:2008} by comparing its predictions with Monte Carlo simulations reported
in the literature \cite{stroobants:1987,veerman:1991}. We will consider all possible
non-uniform phases, namely N, Sm, C and K and will depict the phase diagram the FMF
predicts. There is an interesting aspect about this model that poses a particularly
stringent test on the theory. In Ref.~\cite{stroobants:1987} a window of stability of
the C phase was reported whose existence the authors of Ref.~\cite{veerman:1991}
could not completely settle, although their results pointed to its being a finite
size effect because this window disappears ---being preempted
by a K--- in simulations of very large systems. We will show that our FMF does
indeed confirm this conclusion by showing that either the Sm or the K are always
more stable than the C, although the difference in free energy is rather small
---what justifies its observation in small systems. We will also compare the
resulting equations of state for the N, Sm and K phases with those obtained from
the Monte Carlo simulations of Ref.~\cite{veerman:1991} and conclude that the performance
of our functional is almost perfect in the description of highly non-uniform phases,
even improving on the free-volume description of the K phase.  

\section{Fundamental measure density functional}
\label{cylinders}

In \cite{martinez-raton:2008} we obtained a fundamental-measure density functional
for mixtures of parallel hard cylinders, so we will just gather here the formulae, specialized
for the case of a one-component fluid. The functional is constructed out of
the one for two-dimensional hard disk. There are two versions for the latter:
Rosenfeld's original version \cite{rosenfeld:1990}, and the version of
Tarazona and Rosenfeld \cite{tarazona:1997}. The former has some important
drawbacks, for instance, the low density limit of the functional is only
approximate. That of Tarazona and Rosenfeld recovers the exact result in this
limit. On the other hand, the former is easier to implement than the latter,
because it is expressible in terms of one-particle-weighted densities, while
that of Tarazona and Rosenfeld contains a two-particle-weighted density.
Nevertheless both are amenable to numerical treatment and we will explore
the results of both. So the formulae presented here will describe the
implementation of the two versions for the functional of parallel hard cylinders.

Irrespective of the version we are using, the free-energy density functional
can always be written
\begin{equation}
\beta\mathcal{F}[\rho]=\beta\mathcal{F}_{\rm id}[\rho]
+\beta\mathcal{F}_{\rm ex}[\rho],
\label{total}
\end{equation}
where $\beta$ is the inverse temperature in units of the Boltzmann constant,
\begin{equation}
\beta\mathcal{F}_{\rm id}[\rho]=\int d{\bf r}\int dz\,\rho({\bf r},z)
\left[\ln\mathcal{V}\rho({\bf r},z)-1\right]
\end{equation}
is the functional of the ideal gas ($\mathcal{V}$ is the thermal volume,
irrelevant for the phase behavior), and $\beta\mathcal{F}_{\rm ex}[\rho]$
is the excess free energy due to interactions. We are using the notation
${\bf r}=(x,y)$ for vectors perpendicular to the cylinders axes.
Fundamental-measure functionals are expressed in terms of an excess
free-energy density $\Phi({\bf r},z)$, such that
\begin{equation}
\beta\mathcal{F}_{\rm ex}[\rho]=\int d{\bf r}\int dz\,\Phi({\bf r},z).
\end{equation}
This free-energy density can be given as a function of a set of weighted
densities. The whole set of them can be written in terms of the two
densities
\begin{eqnarray}
\rho_0({\bf r},z) &=& \frac{1}{2}\left[\rho({\bf r},z+L/2)+
\rho({\bf r},z-L/2)\right], \\
\rho_1({\bf r},z) &=& \int_{z-L/2}^{z+L/2}\rho({\bf r},t)\,dt.
\end{eqnarray}
Common to both versions are the weighted densities
\begin{eqnarray}
n_0({\bf r},z) &=& \frac{1}{2\pi R}\int\limits_{|{\bf R}|=R}
\rho_0({\bf r}+{\bf R},z)\,d{\bf R}, \label{eq:n0} \\
n_1({\bf r},z) &=& \frac{1}{2\pi R}\int\limits_{|{\bf R}|=R}
\rho_1({\bf r}+{\bf R},z)\,d{\bf R}, \label{eq:n1} \\
n_2({\bf r},z) &=& \int\limits_{|{\bf R}|\le R}
\rho_0({\bf r}+{\bf R},z)\,d{\bf R}, \label{eq:n2} \\
n_3({\bf r},z) &=& \int\limits_{|{\bf R}|\le R}
\rho_1({\bf r}+{\bf R},z)\,d{\bf R}. \label{eq:n3}
\end{eqnarray}
For Rosenfeld's original version \cite{rosenfeld:1990} there are
also two vector densities, namely
\begin{eqnarray}
{\bf v}_1({\bf r},z) &=& \frac{1}{2\pi R^2}\int\limits_{|{\bf R}|=R}
\rho_0({\bf r}+{\bf R},z){\bf R}\,d{\bf R}, \\
{\bf v}_2({\bf r},z) &=& \frac{1}{2\pi R^2}\int\limits_{|{\bf R}|=R}
\rho_1({\bf r}+{\bf R},z){\bf R}\,d{\bf R},
\end{eqnarray}
and the expression for the excess free-energy density is
\begin{equation}
\Phi_{\rm Ros}=-n_0\ln(1-n_3)+\frac{n_1n_2+2\pi R^2(n_0n_1-
{\bf v}_1\cdot{\bf v}_2)}{1-n_3}+\pi R^2n_2\,\frac{n_1^2-{\bf v}_2^2}{(1-n_3)^2}.
\label{eq:phiR}
\end{equation}
For Tarazona-Rosenfeld's version \cite{tarazona:1997} there are
also two two-particle-weighted densities, namely
\begin{eqnarray}
N_1({\bf r},z) &=& \hspace*{-4mm}\int\limits_{|{\bf R}_1|=R_1}
\hspace*{-4mm}d{\bf R}_1 \hspace*{-3mm} \int\limits_{|{\bf R}_2|=R_2}
\hspace*{-4mm}d{\bf R}_2\,
\left[\rho_1({\bf r}+{\bf R}_1,z)\rho_0({\bf r}+{\bf R}_2,z) 
+\rho_0({\bf r}+{\bf R}_1,z)\rho_1({\bf r}+{\bf R}_2,z)\right]
\nonumber \\
&&\times K\left(\frac{|{\bf R}_1-{\bf R}_2|}{2R}\right), \\
N_2({\bf r},z) &=& \hspace*{-4mm}\int\limits_{|{\bf R}_1|=R_1}
\hspace*{-4mm}d{\bf R}_1
\hspace*{-3mm}
\int\limits_{|{\bf R}_2|=R_2}
\hspace*{-4mm}d{\bf R}_2\,\rho_1({\bf r}+{\bf R}_1,z)
\rho_1({\bf r}+{\bf R}_2,z)K\left(\frac{|{\bf R}_1-{\bf R}_2|}{2R}\right),
\end{eqnarray}
where
\begin{equation}
K(x)=\frac{x}{\pi}\sqrt{1-x^2}\sin^{-1}x,
\end{equation}
and the expression for the excess free-energy density is
\begin{equation}
\Phi=-n_0\ln(1-n_3)+\frac{n_1n_2+N_1}{1-n_3}+\frac{n_2N_2}{(1-n_3)^2}.
\label{eq:phiTR}
\end{equation}

\section{Phase behavior}
\label{results}

\begin{figure}
\epsfig{file=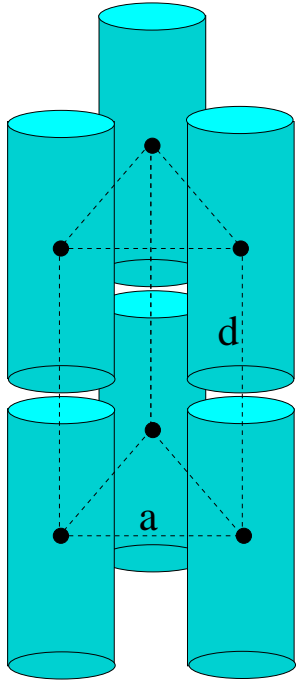,width=1.5in}
\caption{(Color online) Triangular (AAA) crystal. The lattice parameters $a$ and $d$
are shown in the figure.}
\label{lattice}
\end{figure}

The Euler-Lagrange equation
\begin{equation}
\frac{\delta\beta\mathcal{F}}{\delta\rho({\bf r},z)}=\beta\mu,
\label{eq:E-L}
\end{equation}
provides the equilibrium density
for the system when there is no external field and the chemical potential
is fixed to $\mu$ (equivalently, when the mean density is fixed to the
value $\rho$ corresponding to that chemical potential).
Expected phases are nematic (no
spatial ordering), smectic (one-dimensional layering of particles),
columnar (two-dimensional odering of liquid columns) and crystal (a
combination of both orderings). These are the phases shown in the simulations
of Veerman and Frenkel \cite{veerman:1991}. Quite as expected, columnar
phase is a triangular ordering of columns and crystal phase is a piling up
of such triangular lattices, i.e.\ what is commonly referred to as an AAA
crystal (see Fig.~\ref{lattice}).

A direct solution to
(\ref{eq:E-L}) is numerically unfeasible so, as it is customary, we have
resorted to a variational method. Thus, in order to account for all the
above phases in our density functional description in a unified simple way,
we have chosen the parametrization
\begin{eqnarray}
\rho({\bf r})&=&\rho V_{\rm{cell}}^{(D)}\chi_{\perp}({\bf r})
\chi_{\parallel}(z), 
\label{profiles}
\end{eqnarray}
where $\rho$ is the mean density (number of particles per unit volume) and
\begin{eqnarray}
\chi_{\perp}({\bf r})&=&
\frac{\alpha_{\perp}}{\pi}\sum_{\bf k} 
\exp\left[-\alpha_{\perp}\left({\bf r}-{\bf R}_{\bf k}\right)^2\right], \\
\chi_{\parallel}(z)&=&\left(\frac{\alpha_{\parallel}}{\pi}
\right)^{1/2}\sum_{k_3}\exp\left[-\alpha_{\parallel}(z-k_3d)^2\right].
\end{eqnarray}  
The parameter $V_{\rm{cell}}^{(D)}$ is defined as the $D$-dimensional
volume of the unit cell of the corresponding phase ($D=1$ smectic, $D=2$
columnar, $D=3$ crystal). Its values are
\begin{equation}
V^{(1)}_{\rm cell}=d, \qquad
V^{(2)}_{\rm cell}=\sqrt{3}a^2/2, \qquad
V^{(3)}_{\rm cell}=d\sqrt{3}a^2/2,
\label{eq:Vcell}
\end{equation}
$d$ being the layer spacing along the Z direction and $a$ the lattice parameter
of the triangular lattice on the XY plane (see Fig.~\ref{lattice}).
Finally, ${\bf R}_{\bf k}=k_1{\bf a}_1+k_2{\bf a}_2$ ($k_1,k_2\in\mathbb{Z}$),
with $\displaystyle{{\bf a}_n=\frac{a}{2}\left(\sqrt{3},
(-1)^n\right)}$ the vectors defining the two-dimensional
triangular lattice.
In Appendix~\ref{explicit} we give explicit expressions for the weighted
densities evaluated with the density profile (\ref{profiles}).
%
%

When Eq.~(\ref{eq:E-L}), using the parametrization (\ref{profiles}), leads
to a solution with
$\alpha_{\parallel}=\alpha_{\perp}=0$, the equilibrium phase is a nematic;
a smectic is the equilibrium phase if $\alpha_{\parallel}\ne 0$ and
$\alpha_{\perp}=0$; it is a columnar if $\alpha_{\parallel}=0$ and
$\alpha_{\perp}\ne 0$; and a crystal if both $\alpha_{\parallel}\ne 0$
and $\alpha_{\perp}\ne 0$. For the crystal 
phase $1-\rho V_{\rm{cell}}^{(3)}=\nu$ provides the fraction of vacancies. 

\subsection{Nematic phase}

\begin{figure}
\epsfig{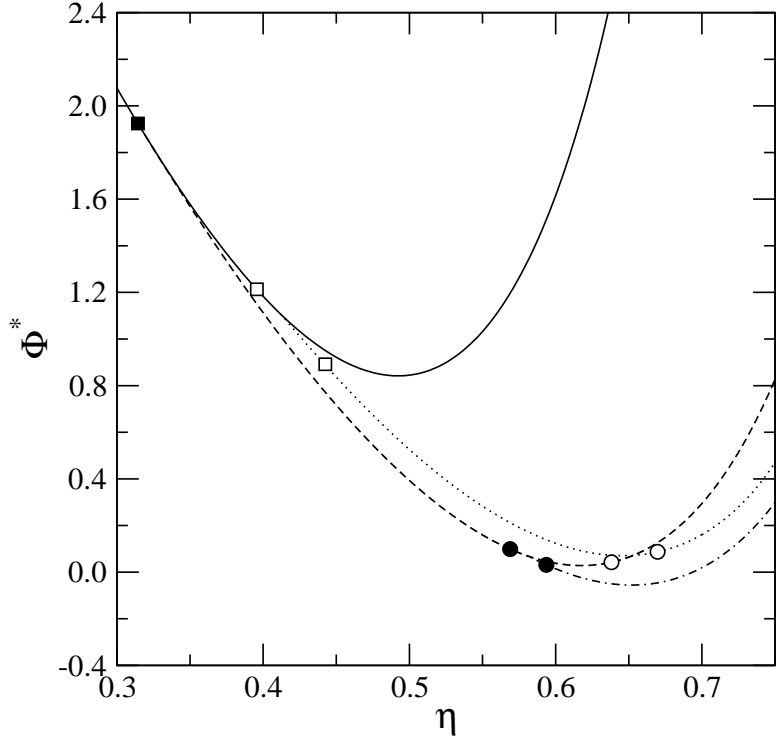}
\caption{Free-energy densities $\Phi^*=\overline{\Phi}-\overline{\Phi}_0-a_1\eta-a_2$ 
(with $a_1=4.8463$ and $a_2=-2.0555$ chosen so as to amplify the differences between 
the different free-energy branches) vs.\ packing fraction $\eta$ for the 
nematic (solid line), smectic (dashed line), columnar (dotted line) 
and crystal (dash-dotted line) phases. The N-Sm bifurcation point is shown by a
filled square. The nematic-columnar and smectic-columnar coexisting packing
fractions are marked with open squares and open circles, respectively. The columnar
phase is metastable and hence so are these two phase transitions. Before the
smectic changes into a columnar the crystal becomes more stable. The smectic-crystal
phase transition is marked with full circles.} 
\label{energy}
\end{figure}

When $\alpha_{\parallel}=\alpha_{\perp}=0$ in (\ref{profiles}), both
(\ref{eq:phiR}) and (\ref{eq:phiTR}) provide the same free-energy density, namely
\begin{equation}
\overline{\Phi}\equiv\frac{\beta Fv}{V}=\overline{\Phi}_0+
\eta(\ln y+3y+y^2),
\label{eq:freeenergy}
\end{equation}
where $\overline{\Phi}_0=\eta\ln(\mathcal{V}/v)-\eta$,
a linear term irrelevant for phase behavior, $\eta=\rho v$
is the packing fraction, $v=\pi R^2L$ is the volume of a cylinder,
and $y=\eta/(1-\eta)$. This free-energy density is plotted in Fig.~\ref{energy}.

{}From (\ref{eq:freeenergy}) the equation of state is readily obtained as
\begin{equation}
\beta pv=y+3y^2+2y^3=\eta\,\frac{1+\eta}{(1-\eta)^3},
\end{equation}
the same equation of state as that of parallel hard cubes \cite{martinez-raton:1999}.

The structure factor can also be obtained from the
relationship $S(q,q_z)=[1-\rho\hat c(q,q_z)]^{-1}$,
where $\hat c(q,q_z)$ is the Fourier transform of the direct correlation function
of the uniform fluid. Its expression was given in Ref.~\cite{martinez-raton:2008}
[cf.\ Eqs.~(39)--(43) and Appendix~B]. Specializing to the one-component fluid
and taking into account that
\begin{eqnarray}
\int d{\bf q}\,\delta(u-r) &=& 2\pi u\Psi_0(qu), \\
\int d{\bf q}\,\Theta(u-r) &=& \pi u^2\Psi_1(qu), \\
\int dq_z\,\Theta(u/2-|z|) &=& u\Psi_2(q_zu/2),
\end{eqnarray}
where ${\bf q}=(q_x,q_y)$, $q=|{\bf q}|$, $r=|{\bf r}|$, $\Psi_0(x)=J_0(x)$,
$\psi_1(x)=2J_1(x)/x$ and $\Psi_2(x)=\sin x/x$, $J_0(x)$ and $J_1(x)$ being
the zeroth and first order Bessel functions, respectively, we obtain, from
the Tarazona-Rosenfeld functional (\ref{eq:phiTR}), the
following expression for the inverse structure factor
\begin{equation}
\begin{split}
S(q,q_z)^{-1} =& 1+8y\Psi_1(2q^*)\Psi_2(2q_z^*)+
4y^2\left[2\Psi_0(q^*)\Psi_1(q^*)\Psi_2(2q_z^*)+\Psi_1(2q^*)\Psi_2(q_z^*)^2\right] \\
&+2y^2(1+2y)\left[2\Psi_0(q^*)\Psi_1(q^*)\Psi_2(q_z^*)^2
+\Psi_1(q^*)^2\Psi_2(2q_z^*)\right] \\
&+y^2(1+6y+6y^2)\Psi_1(q^*)^2\Psi_2(q_z^*)^2,
\end{split}
\label{eq:structure}
\end{equation}
where $q^*=Rq$ and $q_z^*=Lq_z/2$.

\subsection{Smectic phase}

When we set $\alpha_{\perp}=0$ in (\ref{profiles}) and substitute this density
profile into either (\ref{eq:phiR}) or (\ref{eq:phiTR}), both yield the same
expression
\begin{equation}
\Phi(z)=n_0(z)\left\{-\ln\left[1-n_3(z)\right]+\frac{3n_3(z)}{1-n_3(z)}+
\frac{n_3(z)^2}{\left[1-n_3(z)\right]^2}\right\},
\end{equation}
with
\begin{eqnarray}
n_0(z)&=&\frac{1}{2}\left[\rho(z-L/2)+\rho(z+L/2)\right],\\
n_3(z)&=&\pi R^2\int_{z-L/2}^{z+L/2}dz'\rho(z').   
\end{eqnarray}
So both theories predict the same nematic-smectic transition.

Solving Eq.~(\ref{eq:E-L}), a solution with $\alpha_{\parallel}\ne 0$ is
found for every $\eta>\eta_{\rm N-Sm}\approx 0.31$ (also plotted in
Fig.~\ref{energy}). The value of $\alpha_{\parallel}$
approaches zero as $\eta$ approaches this value from above. On the other hand,
the free-energy density for this smectic phase is tangent to that of the
nematic one (see Fig.~\ref{energy}), so the transition is continuous. This
being so, we can obtain a more accurate value of $\eta_{\rm N-Sm}$ as the
smallest $\eta$ at which the structure factor (\ref{eq:structure}) diverges at
some wave vector ${\bf q}={\bf 0}$, $q_z\ne 0$. Specializing (\ref{eq:structure})
for such a wave vector we find
\begin{equation}
S(0,q_z)^{-1}= 1+2y(4+5y+2y^2)\Psi_2(2q_z^*)+
y^2(9+14y+6y^2)\Psi_2(q_z^*)^2.
\label{eq:structure-Sm}
\end{equation}
The smallest $\eta$ for which the right-hand side of (\ref{eq:structure-Sm})
vanishes at a $q_z^*$ is $\eta_{\rm N-Sm}=0.3143$, and the value
of $q_z^*$ at which it happens corresponds to a smectic period $d/L=\pi/q^*_z
=1.3015$.

\subsection{Columnar phase}

At packing fraction $\eta^*_{\rm N-C}=0.4369$ the nematic loses stability
against columnar ordering.
This value is determined from the divergence of the structure factor
(\ref{eq:structure}) at a wave vector ${\bf q}\ne{\bf 0}$, $q_z=0$, which,
for the Tarazona-Rosenfeld functional
(\ref{eq:phiTR}), is given by
\begin{equation}
S(q,0)^{-1} = 1+ 4y(2+y)\Psi_1(2q^*)+4y^2(3+2y)\Psi_0(q^*)
\Psi_1(q^*) +y^2(3+10y+6y^2)\Psi_1^2(q^*).
\label{eq:structure-C}
\end{equation}
In this case, however, the columnar free
energy is not tangent to the nematic one, so the transition is first order. We
can determine the N-C coexistence by the usual double tangent construction. This
yields the $\eta_{\rm N}=0.3957$ and $\eta_{\rm C}=0.4425$ as the coexisting packing
fractions of the nematic and the columnar phases, respectively (see Fig.~\ref{energy}).
At the latter, the lattice parameter is $a/R=2.4744$.

We can see here an important difference between this version of the functional and
that based on Rosenfeld's original approximation, Eq.~(\ref{eq:phiR}). The latter leads
to the following equation for the inverse structure factor
\begin{eqnarray}
S(q,0)^{-1} &=& 1+ 
2y(2+y)\left(\Psi_0^2(q^*)-|\boldsymbol{\Psi}_0(q^*)|^2\right)
+2y(2+7y+4y^2)\Psi_0(q^*)\Psi_1(q^*)\nonumber\\ 
&&+y^2(3+10y+6y^2)\Psi_1^2(q^*),
\label{yasha_v}
\end{eqnarray} 
where the new complex vector 
$\displaystyle{\boldsymbol{\Psi}_0(q^*)=iJ_1(q^*){\bf q}/q}$
has been introduced. The value of $\eta^*_{\rm N-C}$ which this approximation
predicts is $\eta^*_{\rm N-C}=0.5599$. If we had to believe this value for the
N-C bifurcation, the columnar free energy would be much too high to be consistent
with the metastable columnar phase found in simulations
\cite{stroobants:1987,veerman:1991}. For this reason,
we have not pursued this version of the funcional anymore.

The columnar free energy is higher than the smectic one up to $\eta=0.6534$, where a
first order Sm-C transition occurs, with coexisting packing fractions $\eta_{\rm Sm}=0.6382$
and $\eta_{\rm C}=0.6697$. However at these packing fractions the equilibrium phase is no more
the smectic but the crystal, thus the columnar phase is always metastable, and in particular
so are the N-C and the Sm-C transitions. All this can be easily visualized in Fig.~\ref{energy}.

\subsection{Crystal phase}

At packing fractions around $\eta\approx 0.58$ a solution to Eq.~(\ref{eq:E-L}) with
$\alpha_{\parallel}\ne 0$ and $\alpha_{\perp}\ne 0$ renders a free energy smaller than
that of the, up to that point stable, smectic phase. The fluid undergoes a first order
Sm-K transition with coexisting packing fractions $\eta_{\rm Sm}=0.5689$ and
$\eta_{\rm K}=0.5936$. The lattice parameters of the coexisting crystal are
$a/R=2.3102$ and $d/L=1.1419$. With these values the fraction of vacancies can be
found to be just a mere 0.3\%. The crystal is the only stable phase for $\eta > \eta_{\rm K}$
up to close packing (see Fig.~\ref{energy}).

\section{Comparison with computer simulations}

Numerical simulations for this fluid were carried out first by Stroobants et al.\
\cite{stroobants:1987} and later by Veerman and Frenkel \cite{veerman:1991}. The
former, made with $900$ cylinders, showed the sequence of stable phases N-Sm-C-K.
The latter confirmed this result but also made simulations with 1080 cylinders
which showed that the columnar phase previously found appeared due to a finite
size effect. Their conclusion was that the columnar phase is always metastable,
but has a free energy very close to that of the smectic phase, so much that the
boundary conditions may artificially render it more stable. Our previous calculations
are fully consistent with this result, as Fig.~\ref{energy} illustrates.

\begin{figure}
\epsfig{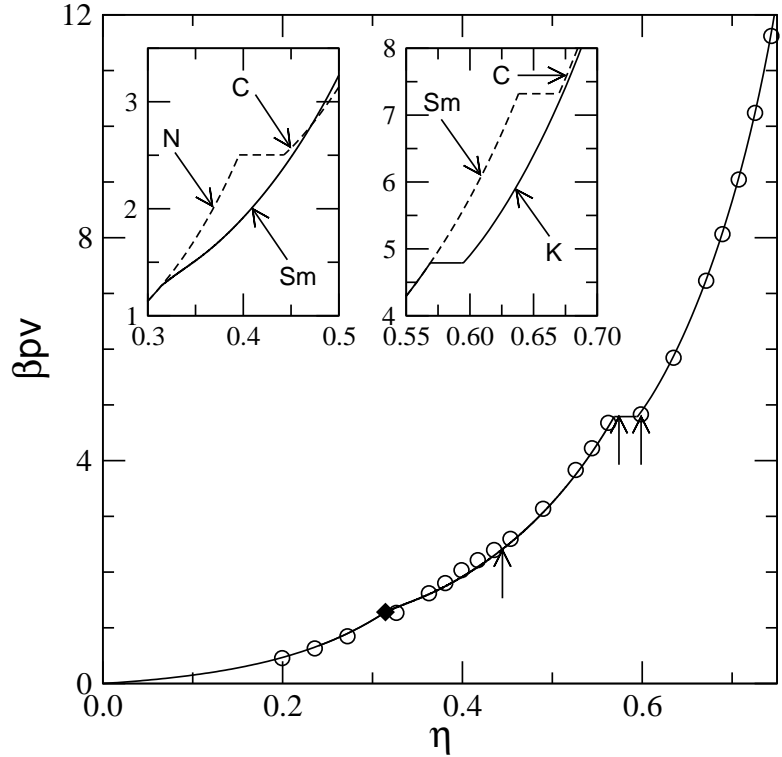}
\caption{Equations of state (reduced pressure vs.\ packing fraction) for all stable
phases obtained from the fundamental measure functional for parallel hard cylinders.
These phases are: nematic (for packing fractions up to the point indicated by a full
rhombus), smectic (from that point up to the discontinuity) and crystal (from the
discontinuity up to close packing). The open circles are the simulation results reported
in Ref.~\cite{veerman:1991}. Arrows mark the nematic-smectic and smectic-cystal phase\
transitions as obtained from those simulations. The two insets show the equations of 
state for the columnar metastable phase in the neighborhood of the nematic-columnar
(left inset) and smectic-columnar (metastable) phase transitions. [Labels stand for 
nematic (N), smectic (Sm), columnar (C) and crystal (K).]
} 
\label{EOS1}
\end{figure}

Besides this first qualitative agreement, we can also perform a more quantitative
comparison with simulations by comparing the equations of state. This is done in
Fig.~\ref{EOS1}. The simulation results are those obtained with the largest system
size \cite{veerman:1991}. The figure shows that the agreement between the numerical
values of the pressure is excellent for all stable phases. The values for the crystal
phase are indistinguishable from the simulations, as it is also the location of
the Sm-K transition.

The only important deviation between theory and simulations
concerns the location of the N-Sm transition. While both, theory and simulation, 
predict that this transition is continuous, the theory predicts that it occurs at    
$\eta=0.3143$ while the simulations yield a value of $\eta=0.443$. This failure
of the theory to predict the location of continuous transitions between  low-density
uniform and non-uniform phases is a fingerprint of FMT. For instance, the FMF of
parallel hard cubes also predicts the same value of $\eta=0.3143$ for the transition
between the fluid and the smectic, columnar and crystal phases (the later being the
stable one) \cite{cuesta:1997b,martinez-raton:2004}, while simulations provide a
value of $\eta=0.49$ for the freezing of this fluid \cite{jagla:1998,groh:2001}.
The reason for this drawback lies in the fact that, by construction, FMFs provide,
in the uniform limit, the SPT equation of state ---which for anisotropic bodies
deviates from the exact result---, while at the same time the
prediction for the nonuniform phases improves significantly
due to the dimensional crossover properties of FMFs \cite{tarazona:2002}.
This discrepancy in the accuracy with which the theory describes both type of
phases leads to inaccurate predictions of the uniform-nonuniform phase transition
points.

We end this section by comparing the EOS for the crystal phase given by the FMF and
that obtained by a cell approximation for the fluid of parallel hard cylinders, which
is derived in Appendix~\ref{sec:cell}. Figure~\ref{cell-theory} shows the results of both
theories as well as the simulation results. As it can be seen, while the FMF results
fit perfectly the simulation points, the cell approximation, although still 
a rather good description, underestimates the EOS. We can also see that, as expected,
both theories converge at high densities, a known result which is a direct consequence
of the dimensional crossover 3D$\to$0D of the FMF \cite{rosenfeld:1996,rosenfeld:1997}. 

\begin{figure}
\epsfig{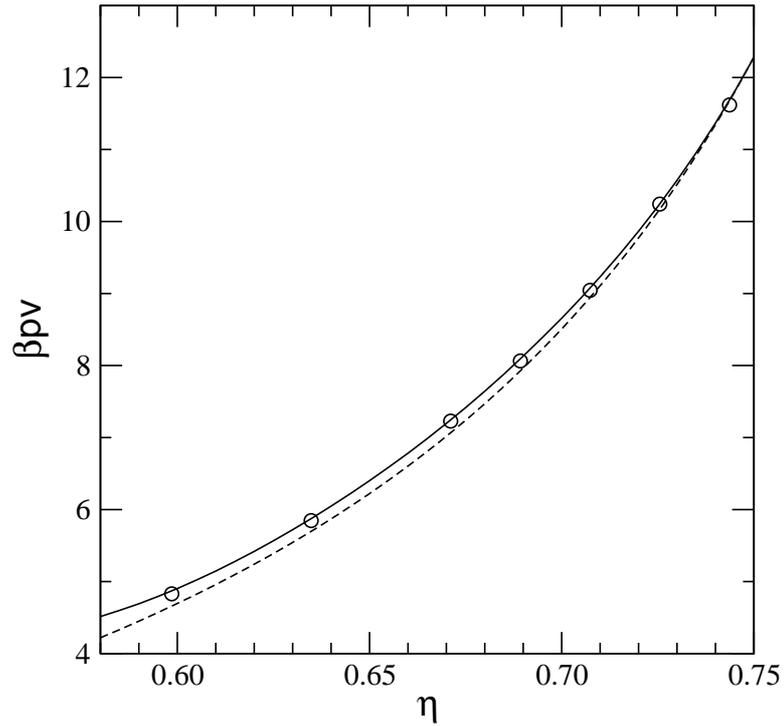}
\caption{Comparison between the equation of state of the crystal phase as obtained from
minimization of the functional (solid line), from the cell approximation (dashed line) and
from simulations \cite{veerman:1991}.}
\label{cell-theory}
\end{figure}

\section{Conclusions}

There are very few examples in the literature in which the same functional describes with
accuracy all inhomogeneous phases of a liquid crystalline fluid. In this article we have
applied a fundamental-measure functional recently proposed for mixtures of parallel hard
cylinders \cite{martinez-raton:2008} to determine the phase behavior of the one-component
fluid. As usual with fundamental-measure-based functionals, the results obtained for the
uniform (nematic) fluid are those provided by scaled particle theory, and so the accuracy
the functional provides for this phase is reasonably good but not perfect. As a consequence,
the predicted nematic-smectic phase transition deviates significantly from the Monte Carlo
simulations of Refs.~\cite{stroobants:1987,veerman:1991}, although the order is correct.
However, the accuracy with which the remaining stable phases, smectic and crystal, are
obtained is excellent, the plots being indistinguishable from the simulation data, even
for the smectic-crystal coexisting densities. Results for the equation of state of the
crystal improve on those obtained by a cell approximation (which we have also reported in
an appendix). Another correct prediction of the theory is that the columnar is only a 
metastable phase, but its free energy is sufficiently close to that of the stable phases
so as to justify the observation of a window of stability of that phase in the oldest
simulations \cite{stroobants:1987} made with the smallest system size, a window that
disappears when the size in increased \cite{veerman:1991}. In summary, the proposed
functional provides excellent results, very similar to those obtained by simulations,
but obtained at a much cheaper price. They also made us confident that its version for
mixture may provide very good results as well.

\acknowledgments

J.\ A.\ Capit\'an acknowledges financial support through a contract from Consejer\'{\i}a
de Educaci\'on of Comunidad de Madrid and Fondo Social Europeo. Y.\ Mart\'{\i}nez-Rat\'on
was supported by a Ram\'on y Cajal research contract. This work is part of research projects
MOSAICO of the Ministerio de Educaci\'on y Ciencia (Spain), and MOSSNOHO of Comunidad 
Aut\'onoma de Madrid (Spain).

\appendix

\section{Explicit expressions for the weighted densities}
\label{explicit}

Insertion of the parametrization (\ref{profiles}) into the expressions for
the weighted densities (\ref{eq:n0})--(\ref{eq:n3}) leads to the formulae
\begin{eqnarray}
n_0({\bf r},z)= \rho V_{\rm{cell}}^{(D)} 
Q_{\phantom{i}\perp}^{(D)}({\bf r}) P_{\phantom{i}\parallel}^{(D)}(z),\\
n_1({\bf r},z)= \rho V_{\rm{cell}}^{(D)} 
Q_{\phantom{i}\perp}^{(D)}({\bf r}) H_{\phantom{i}\parallel}^{(D)}(z),\\
n_2({\bf r},z)= \rho V_{\rm{cell}}^{(D)} 
T_{\phantom{i}\perp}^{(D)}({\bf r}) P_{\phantom{i}\parallel}^{(D)}(z),\\
n_3({\bf r},z)= \rho V_{\rm{cell}}^{(D)} 
T_{\phantom{i}\perp}^{(D)}({\bf r}) H_{\phantom{i}\parallel}^{(D)}(z),
\end{eqnarray}
where $V_{\rm{cell}}^{(D)}$ is defined in Eq.~(\ref{eq:Vcell}).
The functions are given in terms of 
\begin{eqnarray}
g_{\alpha}(x)&=&\left(\frac{\alpha}{\pi}\right)^{1/2} e^{-\alpha x^2},\qquad
e_{\alpha}(x)=\frac{1}{2}\,{\rm erf}(\sqrt{\alpha}x),
\end{eqnarray}
${\rm erf}(x)$ being the standard error function. To be precise,
\begin{eqnarray}
Q_{\phantom{i}\perp}^{(1)}({\bf r}) &=& 1, \\
Q_{\phantom{i}\perp}^{(2)}({\bf r}) &=&
Q_{\phantom{i}\perp}^{(3)}({\bf r}) = 
g_{\alpha_{\perp}}(R) \sum_{{\bf k}} 
g_{\alpha_{\perp}}(|{\bf r} - {\bf R}_{\bf k}|)
I_0(2R\alpha_{\perp}|{\bf r} - {\bf R}_{\bf k}|),
\end{eqnarray}
where $I_0$ stands for the zeroth-order modified Bessel function of the first
kind. The rest of the expressions are similar;
\begin{eqnarray}
T_{\phantom{i}\perp}^{(1)}({\bf r})&=& \pi R^2,\\
T_{\phantom{i}\perp}^{(2)}({\bf r})&=& 
T_{\phantom{i}\perp}^{(3)}({\bf r})= 
2\pi \sum_{\bf k} g_{\alpha_{\perp}}(|{\bf r} - {\bf R}_{\bf k}|)
\int_0^{R} dt\,t\,g_{\alpha_{\perp}}(t)\,
I_0(2t\alpha_{\perp}|{\bf r} - {\bf R}_{\bf k}|),\\
P_{\phantom{i}\parallel}^{(2)}(z)&=& 1,\\
P_{\phantom{i}\parallel}^{(1)}(z)&=&
P_{\phantom{i}\parallel}^{(3)}(z)=\frac{1}{2} 
\sum_{k} [ g_{\alpha_{\parallel}}(z-kd+L/2) +
g_{\alpha_{\parallel}}(z-kd-L/2) ],\\
H_{\phantom{i}\parallel}^{(2)}(z)&=& L, \\
H_{\phantom{i}\parallel}^{(1)}(z)&=& 
H_{\phantom{i}\parallel}^{(3)}(z) =
\sum_{k} [ e_{\alpha_{\parallel}}(z-kd+L/2) -
e_{\alpha_{\parallel}}(z-kd-L/2) ].
\end{eqnarray}

As for the two-particle weighted densities, after a lengthy
calculation (see Ref.~\cite{martinez-raton:2008} for some details) $N_1$ can
be expressed as
\begin{eqnarray}
N_1({\bf r},z)=2(\rho V_{\rm{cell}}^{(D)})^2
P_{\phantom{i}\parallel}^{(D)}(z) H_{\phantom{i}\parallel}^{(D)}(z)
J_{\phantom{i}\perp}^{(D)}({\bf r}),
\end{eqnarray}
with the functions $P_{\phantom{i}\parallel}^{(D)}$ and $H_{\phantom{i}\parallel}^{(D)}$
defined above. The radial contribution is 
\begin{eqnarray}
J_{\phantom{i}\perp}^{(1)}({\bf r})  &=& \pi R^2, \\
J_{\phantom{i}\perp}^{(2)}({\bf r}) &=&
J_{\phantom{i}\perp}^{(3)}({\bf r}) =
\left(\frac{\alpha_{\perp}}{\pi}\right)^2\!R^2e^{-2R^2\alpha_{\perp}}
\!\!\sum_{{\bf k}_1,{\bf k}_2}\!
e^{-\alpha_{\perp}[({\bf r}-{\bf R}_{{\bf k}_1})^2+({\bf r}-{\bf R}_{{\bf k}_2})^2]}
\nonumber \\
&&\times
\int_0^{\pi}dt\, t \sin t\, I_0[B_{{\bf k}_1,{\bf k}_2}(t,{\bf r})],
\end{eqnarray}
where
\begin{eqnarray}
B_{{\bf k}_1,{\bf k}_2}(t,{\bf r})=
2R\alpha_{\perp}\sqrt{\zeta_{{\bf k}_1}^2+\zeta_{{\bf k}_2}^2+2\zeta_{{\bf k}_1}
\zeta_{{\bf k}_2}\cos(t+\psi_{{\bf k}_2}-\psi_{{\bf k}_1})},
\end{eqnarray}
denoting ${\bf r}-{\bf R}_{{\bf k}_{\nu}}=\zeta_{{\bf k}_{\nu}}(
\cos\psi_{{\bf k}_{\nu}},\sin\psi_{{\bf k}_{\nu}})$, with $\nu=1,2$.
Finally, $N_2$, is given by
\begin{eqnarray}
N_2({\bf r},z)=[\rho V_{\rm{cell}}^{(D)}
H_{\phantom{i}\parallel}^{(D)}(z)]^2
J_{\phantom{i}\perp}^{(D)}({\bf r}).
\end{eqnarray}

\section{Cell approximation for the crystal phase of parallel hard cylinders}
\label{sec:cell}

\begin{figure}
\epsfig{file=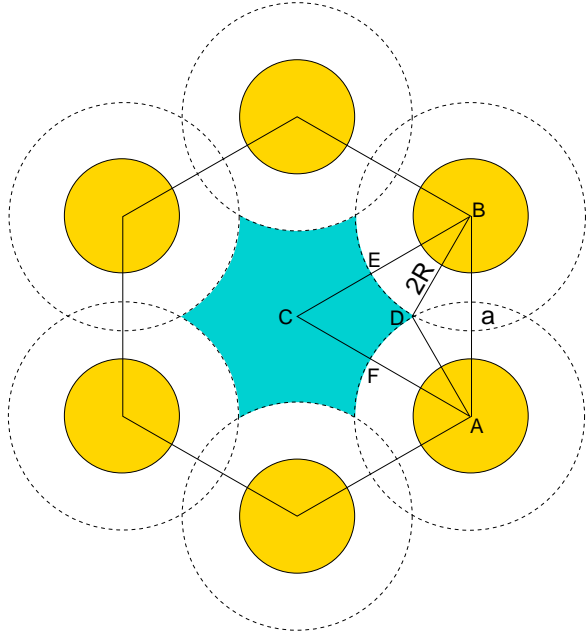,width=3.in}
\caption{(Color online) Sketch of the triangular lattice of period $a$. The free
region of area $A_{\rm{free}}$ within which one particle can move appears colored.
One sixth of this area can be obtained substracting from the area of the triangle
ABC those of the triangle ABD and of the sectors BED and ADF.}
\label{fig:sketch}
\end{figure}

This section is devoted to obtain a cell approximation for the free energy per 
particle of the crystal phase of parallel hard cylinders. To this aim we first
calculate the free volume available to one particle moving in an cell defined 
by the first nearest neighbours: a prism with hexagonal base composed by six
triangular cells of period $a$ (see a sketch in Figure~\ref{fig:sketch}) and
height equal to $2d$. 
Six hard disks (the cylinder sections) of radii $R$ are fixed at the vertexes 
of the hexagon while a seventh one is allowed to move within this cell, with the
only constraint of not overlapping the other six disks (which of course do not
overlap themselves). Simple geometric considerations lead, for the area
accessible to the center of mass of the seventh disk, to the formula
\begin{eqnarray}
A_{\rm{free}}=24R^2\left[\sqrt{3}x^2+\cos^{-1}x -x\sqrt{1-x^2}-
\frac{\pi}{3}\right],
\end{eqnarray}
where $x=a/4R$. The free volume of this  cell is simply 
$V_{\rm{free}}=2A_{\rm{free}}L(y-1)$ with $y=d/L$. 
If we fix the mean packing fraction of the crystal, 
the variables $x$ and $y$ are related through the equation
$\eta=v/V^{(3)}_{\rm{cell}}$, where $v=\pi R^2L$ and $V^{(3)}_{\rm{cell}}$
is defined in (\ref{eq:Vcell}), are the particle and cell volumes respectively. 
Thus $y=\pi/8\sqrt{3}\eta x^2$. 

The cell theory approximates the free energy per particles as 
\begin{eqnarray}
\varphi=
-\ln\left( \frac{V_{\rm{free}}}{\mathcal{V}}\right),
\end{eqnarray} 
with $\mathcal{V}$ the thermal volume of the system, which in our case is
\begin{eqnarray}
\varphi=\ln \left(\frac{\pi\mathcal{V}}{48v}\right)
-\ln\left(\sqrt{3}x^2+\cos^{-1} x -x \sqrt{1-x^2}-
\frac{\pi}{3}\right)-\ln\left(\frac{\pi}{8\sqrt{3}\eta x^2}-1\right).
\label{cell}
\end{eqnarray}
Once the mean packing fraction is fixed the free-energy (\ref{cell}) 
must be minimized with respect to $x$ with the constraint $x\geq 1/2$ 
($x=1/2$ represents the close packed limit), and then the pressure
is obtained as $\displaystyle{\beta P v=\eta^2\frac{\partial \varphi}{
\partial\eta}}$, with the result
\begin{eqnarray}
\beta Pv=\frac{\eta}{1-4x_0^2\eta/\eta_{\rm{cp}}},
\end{eqnarray}
$\eta_{\rm{cp}}=\pi/\sqrt{12}$ being the value of $\eta$ at close packing,
and $x_0$ the solution to the equation 
\begin{eqnarray}
\frac{\eta_{cp}}{4\eta x^2}\left(\cos^{-1} x-\frac{\pi}{3}\right)+
x\left(\sqrt{3}x-\sqrt{1-x^2}\right)=0.
\end{eqnarray}

\bibliography{liquid}
 
\end{document}